\documentclass[prd,superscriptaddress,floatfix,preprintnumbers,altaffilletter,onecolumn,nofootinbib]{revtex4}
\usepackage{graphicx,url,amssymb,amsmath,longtable,rotating,color,units, 
wasysym,subfigure,epsfig,cancel,placeins}
\usepackage[mathscr]{euscript}
\usepackage{color}
\usepackage{verbatim}
\begin{document}

\newcommand{\Hcurl}{\mathscr{H}}
\newcommand{\Fcurl}{\mathscr{F}}
\newcommand*{\rom}[1]{\uppercase\expandafter{\romannumeral #1\relax}}

\pagenumbering{arabic}

\title{Statistical properties of astrophysical gravitational-wave backgrounds}

\author{Duncan Meacher}
\email{Duncan.Meacher@oca.eu}
\affiliation{UMR ARTEMIS, CNRS, University of Nice Sophia-Antipolis, Observatoire de la C$\hat{o}$te d'Azur, BP 4229, 06304, Nice Cedex 4, France}

\author{Eric Thrane}
\affiliation{LIGO Laboratory, California Institute of Technology, MS 100-36, Pasadena, California 91125, USA}

\author{Tania Regimbau}
\affiliation{UMR ARTEMIS, CNRS, University of Nice Sophia-Antipolis, Observatoire de la C$\hat{o}$te d'Azur, BP 4229, 06304, Nice Cedex 4, France}

\date{\today}

\begin{abstract}
We investigate how a stochastic gravitational wave background, produced from a discrete set of astrophysical sources, differs from an idealised model consisting of an isotropic, unpolarised, and Gaussian background. 
We focus, in particular, on the different signatures produced from these two cases, as observed in a cross-correlation search. 
We show that averaged over many realisations of an astrophysical background, the cross-correlation measurement of an astrophysical background is identical to that of an idealised background. 
However, any one realisation of an astrophysical background can produce a different signature. 
Using a model consisting of an ensemble of binary neutron star coalescences, we quantify the typical difference between the signal from individual realisations of the astrophysical background and the idealised case. 
For advanced detectors, we find that, using a cross-correlation analysis, astrophysical backgrounds from many discrete sources are probably indistinguishable from an idealised background.
\end{abstract}

\maketitle

\section{Introduction}\label{intro}
One of the science goals of second-generation gravitational-wave detectors such as Advanced LIGO~\cite{aligo} and Virgo~\cite{virgo} is to detect a stochastic gravitational-wave background. 
A stochastic background arises from the superposition of many gravitational-wave sources, each of which cannot be individually resolved~\cite{Allen97,2000PhR...331..283M}.
A stochastic background can be created in the early universe following inflation~\cite{grishchuk,starob,eastherlim,peloso}, during a phase transition~\cite{2000PhR...331..283M}, or from cosmic strings~\cite{caldwellallen,DV1,DV2,cosmstrpaper,olmez1,olmez2} to name a few scenarios.
Less speculative astrophysical stochastic backgrounds are expected to arise from more vanilla objects such as compact binaries~\cite{phinney,kosenko,zhu,StochCBC,RegPacCBC}, neutron stars~\cite{cutler,RegMan,rosado_ns,howell2010,marassi,RegPac,owen,barmodes2,barmodes3}, core collapse supernovae~\cite{firststars,howell2004,buonanno2005,marassi2009}, white dwarf binaries~\cite{phinney_whitedwarfs} and super-massive black hole binaries~\cite{wyithe,jaffe,enoki}.

A stochastic background can be described in terms of its energy density spectrum $\Omega_\text{gw}(f)$, which is the fractional contribution of the energy density in gravitational waves relative to the total energy density needed to close the universe~\cite{Allen97}:

\begin{equation}
\Omega_\text{gw}(f) = \frac{1}{\rho_c}\frac{d\rho_\text{gw}}{d\ln f} .
\end{equation}

\noindent Here $\rho_c$ is the critical energy density of the universe and $d \rho_\text{gw}$ is the gravitational-wave energy density between $f$ and $f+df$.
Typically, searches for a stochastic background estimate $\Omega_\text{gw}(f)$ using a cross-correlation statistic (see, e.g.,~\cite{Allen97,stoch-S5}), which we denote $\widehat{Y}(f)$.
In~\cite{Allen97}, the estimator $\widehat{Y}(f)$ is derived for the case of an isotropic, unpolarised, and Gaussian background.
While subsequent work has relaxed the assumption of isotropy~\cite{PhysRevD.80.122002,PhysRevLett.107.271102}, it is still typically assumed that the observed background is Gaussian, (see, e.g.,~\cite{stoch-S5}).
However, it is likely that the first detection of a stochastic background will be of a non-Gaussian background of astrophysical origin~\cite{StochCBC}.
Non-Gaussian backgrounds exhibit fluctuations arising from the discrete nature of their composition; no two realisations are exactly the same.

In this paper we investigate how the non-Gaussianity of astrophysical stochastic backgrounds affects cross-correlation measurements of $\Omega_\text{gw}(f)$.
First, we calculate $\langle \widehat{Y}(f)\rangle_{n,h}$, the expectation value of $\widehat{Y}(f)$ in the presence of a non-Gaussian background averaged over both realisations of detector noise and realisations of an astrophysical stochastic background.
The answer, we show, is identical to the case of an isotropic Gaussian background.
Next, we calculate $\langle \widehat{Y}(f)\rangle_n$, the expectation value of $\widehat{Y}(f)$ averaged over realisations of detector noise but considering only {\em a single realisations of an astrophysical background}.
The answer, this time, is different than the case of an isotropic Gaussian background.
By comparing these two calculations, we characterise the signature caused by the discreteness of astrophysical backgrounds.
We proceed to estimate the size of this signature in upcoming observations by advanced detectors.

The remainder of the paper is organised as follows.
In section~\ref{formalism}, we review the procedure for a cross-correlation search for a stochastic background (subsection~\ref{review}), characterise the statistical behaviour of astrophysical backgrounds (subsections~\ref{ensemble} and~\ref{realisation}), and introduce a novel formalism for characterising astrophysical backgrounds.
Then, in section~\ref{numerical}, we present the results of a numerical investigation, which quantifies the statistical fluctuations between different realisations of the stochastic background.
Finally, in section~\ref{conclusions}, we summarise our results and discuss the implication for future gravitational-wave observations.


\section{Formalism}\label{formalism}

\subsection{Cross-correlation searches for a stochastic background}\label{review}
We consider a cross correlation search~\cite{Allen97} using two detectors $i$ and $j$.
The measured strain in detector $i$ is given by

\begin{equation}
s_i(t) = h_i(t) + n_i(t),
\end{equation}

\noindent where $h_i(t)$ is the gravitational-wave strain signal, $n_i(t)$ is the noise, and $t$ is the sample time.
At any given time $t$, there are, we assume, $N_t$ gravitational-wave sources in the universe producing a strain signal.
If the background is very non-Gaussian, $N_t$ may be zero for many values of $t$.
A background where $N_t\gg1$ is quasi-Gaussian.
If $N_t\geq1$, we can write

\begin{equation}\label{eq:hkt}
  h_i(t) = \sum_{k=1}^{N_t} h_{i,k} (t) .
\end{equation}

\noindent (If $N_t=0$, then $h_i(t)=0$.)
Here $h_{i,k}(t)$, the observed strain from the $k^{th}$ gravitational wave source, is implicitly a function of the sky location $\hat\Omega_k$ of the source.
The strain signal can be written as

\begin{equation}\label{eq:h^A}
h_{i,k}(t) = h_{i,k}^A(t) F_{i,k}^A (\hat{\Omega}_k,t) .
\end{equation}

\noindent Here $h_{i,k}^A(t)$ is the Fourier coefficient of a plane-wave metric perturbation in the transverse traceless gauge

\begin{equation}
\begin{split}
h_{ab}(t,\vec x) = \sum_{A=+,\times} \int_{-\infty}^\infty df h_A(f,\hat{\Omega_k}) e^A_{ab}(\hat \Omega_k)\, e^{- 2\pi i f(t-\hat{\Omega_k} \cdot \vec x/c)} ,
\end{split}
\label{e:hab}
\end{equation}

\noindent where $a,b = 1,2,3$ are indices in the transverse plane, $e^A_{ab}(\hat \Omega)$ is the polarisation tensor, $A=+,\times$ is the polarisation, $f$ is frequency, $\vec{x}$ is the position vector of the observer and $c$ is the speed of light.
The $F^A(\hat\Omega,t)$ term in Eq.~\ref{eq:h^A} is the detector response for direction $\hat{\Omega}$ at time $t$ \cite{Allen97}.

We define a strain cross-power estimator in terms of the Fourier transforms of two strain time series

\begin{equation}\label{eq:Y(f)}
\widehat{Y}(f) \equiv Q(f) \sum_t \tilde{s}_i^*(f;t) \tilde{s}_j(f;t) .
\end{equation}

\noindent The sum in Eq.~\ref{eq:Y(f)} is over data segments (typically $\unit[60]{s}$ long; see~\cite{stoch-S5}).
We use $(f;t)$ to denote a Fourier spectrum for a data segment beginning at time $t$, which is in contrast to the sampling time, denoted $(t)$.
Here $Q(f)$ is a filter function chosen such that---if the stochastic background is Gaussian and isotropic---the expectation value of $\widehat{Y}(f)$ is $\Omega_\text{gw}(f)$.
Eq.~\ref{eq:Y(f)} implicitly assumes that the detector noise is stationary.
In the presence of non-stationary detector noise, the equation is modified to weight quiet times as more important than noisy times.
For the sake of simplicity, we present our calculation using the assumption of stationary noise, though, we note that the results are independent of this assumption.

We now consider the expectation value of $\widehat{Y}(f)$ averaging over realisations of detector noise: $\langle\widehat{Y}(f)\rangle_n$.
Here $\langle ... \rangle_n$ denotes the ensemble average over realisations of detector noise

\begin{equation}
\left\langle ... \right\rangle_n \equiv \int dn_i \int dn_j (...) p_n(n_i) \, p_n(n_j) .
\end{equation}

\noindent Here $p_n(n_i)$ and $p_n(n_j)$ are probability density functions describing the noise in detectors $i$ and $j$.
They are typically taken to be normally distributed, and indeed, this assumption is born out in practice; see, e.g.,~\cite{stoch-S5,PhysRevLett.107.271102}.
Here, for the sake of compact notation, we assume that $n_i$ and $n_j$ have the same probability density function $p_n$, though, this assumption can be relaxed without affecting the results.

If the noise in each detector is uncorrelated then $\langle n_i^*(f;t) n_j(f;t) \rangle_n = 0$ and $\langle n_i^*(f;t) h_j(f;t) \rangle_n = 0$ while $\langle h_i^*(f;t) h_j(f;t) \rangle_n \neq 0$ (unless $h_i(f;t)=0$ and/or $h_j(f;t)=0$).
Thus,

\begin{equation}\label{eq:<Y>}
\left\langle \widehat{Y}(f) \right\rangle_n =  Q(f) \sum_t \sum_{k=1}^{N_t} \left\langle \left(h_{i,k}^+(f;t) F_{i,k}^+(t) + h_{i,k}^\times(f;t) F_{i,k}^\times(t)\right)^* \left(h_{j,k}^ + (f;t) F_{j,k}^+(t) + h_{j,k}^\times(f;t) F_{j,k}^\times(t)\right) \right\rangle_n .
\end{equation}

\noindent The parsing of the data into segments is merely a matter of convenience; the sums over $t$ and $k=1...N_t$ are equivalent to a single sum from $k=1...N$ where $N \equiv \sum_t N_t$ (the total number of events that occur during the observation period).
Thus,

\begin{equation}\label{eq:t_to_k}
\sum_t \sum_{k=1}^{N_t} h_{i,k}^A(f;t) F_{i,k}^A(t) h_{j,k}^{A'}(f;t) F_{j,k}^{A'}(t)  = \sum_{k=1}^N h_i^A(f;k) F_i^A(k) h_j^{A'}(f;k) F_j^{A'}(k) .
\end{equation}

\noindent Here $F_i^A(k)$ and $F_j^{A'}(k)$ represent the time-averaged detector response for the $k^\text{th}$ event in detectors $i$ and $j$ respectively.
For most signals of interest for Advanced LIGO and Virgo, the detector response does not vary significantly over the time that the signal is in band, but this need not be the case for lower frequency detectors such as the proposed Einstein Telescope~\cite{ETMDC1}.
Note that since each event is associated with a specific direction $\hat\Omega_k$, $h_i^A(f;k)$ and $F_i^A(k)$ are both implicitly functions of $\hat\Omega_k$.

Combining Eq.~\ref{eq:<Y>} and Eq.~\ref{eq:t_to_k}, it follows that

\begin{equation}\label{eq:<Y>_v2}
\left\langle \widehat{Y}(f) \right\rangle_n = Q(f) \sum_{k=1}^{N} \left\langle \left( h_i^+(f;k) F_i^+(k) + h_i^{\times}(f;k) F_i^{\times}(k)\Big)^* \Big(h_j^+(f;k) F_j^+(k) + h_j^{\times}(f;k) F_j^\times(k)\right) \right\rangle_n .
\end{equation}

\noindent Since each event is associated with a specific direction, the signal for each event at detector $i$ is related to the signal at detector $j$ by a simple phase factor

\begin{equation}\label{eq:i_to_j}
h^A(f;k) \equiv h_j^A(f;k) = h_i^A(f;k) e^{2\pi i f \hat{\Omega}_k \cdot \Delta \vec{x}_k /c} ,
\end{equation}

\noindent where $\Delta \vec{x}_k = \vec{x}_{j,k} -\vec{x}_{i,k}$ is the separation vector between the two detectors at the time of event $k$.  
The vectors $\vec{x}_{i,k}$ and $\vec{x}_{j,k}$ are the positions of detector $i$ and detector $j$ respectively.

Combining Eq.~\ref{eq:<Y>_v2} and Eq.~\ref{eq:i_to_j}, we obtain

\begin{equation}\label{eq:<Y>_v3}
\begin{split}
\left\langle \widehat{Y} (f) \right\rangle_n = & \; Q(f) \sum_{k=1}^{N} \Bigg\langle e^{2\pi i f \hat{\Omega}_k \cdot \Delta \vec{x}_k /c} \Big( \left| h^+(f;k) \right|^2 F_i^+(k) F_j^+(k) + \left|h^\times(f;k) \right|^2 F_i^\times(k) F_j^\times(k) +  \\
& h^{+*}(f;k) h^\times(f;k) F_i^+(k) F_j^\times(k) + h^{\times*}(f;k) h^+(f;k) F_i^\times(k) F_j^+(k) \Big) \Bigg\rangle_n .
\end{split}
\end{equation}

\noindent In the following subsections we explore the consequences of Eq.~\ref{eq:<Y>_v3}.


\subsection{Average over realisations of a stochastic background}\label{ensemble}

In this subsection, we use Eq.~\ref{eq:<Y>_v3} to derive the expectation value of $\widehat{Y}(f)$ averaged over both detector noise {\em and} over realisations of a stochastic background:

\begin{equation}
\begin{split}
\left\langle \widehat{Y}(f) \right\rangle_{n, h} \equiv & \int dn_i \, p_n(n_i) \int dn_j \, p_n(n_j) \int dN \, p_N(N) \int \prod_{k=1}^N \prod_{A=+,\times} dh^A(f;k) \, \int \frac{d^2\hat\Omega_k} {4\pi} \int \frac{dt_k}{t_\text{obs}} \left( p_h\left(h_i^A\left(f;k\right)\right) \, \widehat{Y}(f) \right) .
\end{split}
\end{equation}

Here $p_N$ is the Poisson-distributed probability density function for the number of events occurring during one observing period (typically of duration $\approx\unit[1]{yr}$).
The $p_h$ term is the probability density function for the strain signal from each event at detector $j$ (see Eq.~\ref{eq:i_to_j}).
(In the next subsection, we focus on a stochastic background from binary neutron stars, which allows us to parameterise $p_h$ in terms of sky location $\hat\Omega$, redshift $z$, inclination angle $\iota$, polarisation angle $\psi$, and chirp mass $M_c$.)
The source direction $\hat\Omega_k$ is assumed to be drawn from an isotropic distribution while the burst time $t_k$ is assumed to be drawn from a uniform distribution on $[0,t_\text{obs}]$.

We assume that $p_h$ is the same for the two polarisation states, which follows from rotational invariance.
Thus, we may define average strain power spectral density per event $\mathfrak{H}(f)$

\begin{equation}\label{eq:H'(f)}
  \left\langle \left| h^+(f;k) \right|^2 \right\rangle_{n,h} = \left\langle \left| h^\times(f;k) \right|^2 \right\rangle_{n,h} \equiv \frac{1}{2}\mathfrak{H}(f) ,
\end{equation}

\noindent On average, the strain power spectral density observed during the full analysis is given by
\begin{equation}\label{eq:H(f)}
  H(f) = N \mathfrak{H}(f) .
\end{equation}

\noindent Strain power spectral density and energy density are simply related by:
\begin{equation}
  H(f) = \frac{3 H_0^2}{2\pi^2}\frac{\Omega_\text{gw}(f)}{f^3} .
\end{equation}
where $H_0$ is the Hubble constant.

Individual sources such as compact binaries often emit elliptically polarised gravitational waves.
However, if the probability distributions for the orientation and sky location of individual sources respect rotational and translational invariance, then the average polarisation of an ensemble of sources is zero:

\begin{equation}\label{eq:cross-terms}
\left\langle h^{+*}(f;k) h^\times(f,k) \right\rangle_{n,h} = \left\langle h^{\times*}(f;k) h^+(f,k) \right\rangle_{n,h} = 0 .
\end{equation}

\noindent We further assume that $H^A(f;k)$ and $\hat{\Omega}_k$ are uncorrelated.

Putting everything together, we obtain

\begin{equation}\label{eq:<<Y>>}
  \begin{split}
    \left\langle \widehat{Y}(f) \right\rangle_{n,h} & = Q(f) \frac{\mathfrak{H}(f)}{2} \sum_{k=1}^N \left\langle e^{2\pi i f \hat{\Omega}_k \cdot \Delta \vec{x}_k /c} \left( F_i^+(k) F_j^+(k) + F_i^\times(k) F_j^\times(k) \right) \right\rangle_{n,h} \\
    & = Q(f) \frac{H(f)}{2N} \sum_{k=1}^N \left\langle e^{2\pi i f \hat{\Omega}_k \cdot \Delta \vec{x}_k /c} \left( F_i^+(k) F_j^+(k) + F_i^\times(k) F_j^\times(k) \right) \right\rangle_{n,h} \\
  \end{split}
\end{equation}

\noindent The only random variables left in Eq.~\ref{eq:<<Y>>} are sky location $\hat\Omega_k$ and emission time $t_k$ since $F^A_i(k)$ and $\Delta\vec{x}_k$ are both implicit functions of $t_k$ and $\hat\Omega_k$.
Thus,

\begin{equation}\label{eq:intk}
  \left\langle e^{2\pi i f \hat{\Omega}_k \cdot \Delta \vec{x}_k /c} F_i^A(k) F_j^A(k) \right\rangle_{n,h} = \int \frac{dt_k}{t_\text{obs}} \int \frac{d \hat\Omega_k}{4\pi} \, e^{2\pi i f \hat{\Omega}_k \cdot \Delta \vec{x}_k /c}  F_i^A(k) F_j^A(k) .
\end{equation}

\noindent The double integral over $t_k$ and $\hat\Omega_k$ can be thought of as a single integral over sky position since an isotropic signal observed at time $t_k$ produces a signal which is identical to the one produced at time $t_k'$.
Thus,

\begin{equation}\label{eq:Gamma}
\begin{split}
\left\langle Y(f) \right\rangle_{n,h} & = Q(f) \frac{H(f)}{2 N} N \int \frac{d \hat\Omega_k}{4\pi} \, e^{2\pi i f \hat{\Omega}_k \cdot \Delta \vec{x}_k /c} \left( F_i^+(k) F_j^+(k) +   F_i^\times(k) F_j^\times(k) \right) , \\
& = Q(f) H(f) \Gamma_{ij}(f) ,
\end{split}
\end{equation}

\noindent where $\Gamma_{ij}(f)$ is the overlap reduction function~\cite{Allen97,christensen_prd,locus} :

\begin{equation}\label{eq:<<Y>>_v2}
\Gamma_{ij}(f) \equiv \frac{1}{8\pi} \int d\hat\Omega \, e^{2\pi i f \hat{\Omega} \cdot \Delta \vec{x}_k /c} \left( F_i^+(\hat\Omega) F_j^+(\hat\Omega) + F_i^\times(\hat\Omega) F_j^\times(\hat\Omega) \right) .
\end{equation}

\noindent Here we use the normalisation convention from~\cite{locus}.

The overlap reduction function encodes information about the interference of gravitational-wave signal coming from different directions on the sky.
Each pair of detectors $ij$ has a different overlap reduction function.
It is also common to define a normalised overlap reduction function $\gamma(f)$ defined such that a colocated coaligned pair has $\gamma_{ij}(f=0)=1$.
For identical interferometers with an opening angle $\delta$,

\begin{equation}\label{eq:gamma}
\gamma_{ij}(f) = (5/\sin^2\delta)\Gamma_{ij}(f) .
\end{equation}

The expression for $\Gamma_{ij}(f)$ given in Eq.~\ref{eq:gamma} is equivalent to the value obtained for an isotropic, unpolarised, Gaussian background~\cite{Allen97}.
This implies that, averaging over realisations of astrophysical backgrounds, a standard search for a stochastic background, assuming an isotropic, unpolarised, Gaussian background will yield an unbiased estimate for the $\hat\Omega(f)$, even if the actual background is non-Gaussian, so long as it is on average unpolarised, and on average isotropic.
In the next subsection we investigate the statistical behaviour of individual realisations of a stochastic background.


\subsection{Individual realisations of a stochastic background for compact binary coalescence}\label{realisation}

In this subsection, we study the expectation value of $\widehat{Y}(f)$ for individual realisations of a stochastic background consisting of a finite number of binary neutron star coalescences.
In the transverse traceless (TT) gauge, the strain signal Fourier coefficients can be written as

\begin{align}
h^{+, TT}_k(f) &= h_{0, k}(z) \frac{\left(1+\cos^2\iota_k \right)}{2} f^{-7/6} , \\
h^{\times, TT}_k(f) &= h_{0, k}(z) \cos\iota_k f^{-7/6} ,
\end{align}

\noindent which are related to the polarisations given in Eq.~\ref{eq:h^A} by

\begin{align}
h^{+}(f) &= h^{+, TT}(f) \cos 2\psi + h^{\times, TT}(f) \sin 2\psi, \\
h^{\times}(f) &= - h^{+, TT}(f) \sin 2\psi + h^{+, TT}(f) \cos 2\psi, 
\end{align}

\noindent where $\psi$ is the angle by which the transverse plane is rotated. The amplitude of the signal is given by

\begin{equation}\label{eq:h_0}
h_{0,k}(z) = \sqrt{\frac{5}{24}} \frac{ \left( GM_{c,k} (1+z_k) \right)^{5/6}}{\pi^{2/3} c^{3/2} d_L(z_k)} .
\end{equation}

\noindent Here $d_L(z)$ is the redshift-dependent luminosity distance and $G$ is the gravitational constant.

We can now rewrite Eq.~\ref{eq:<Y>_v3} as

\begin{align}\label{eq:<Y>_v4}
\begin{split}
  \left\langle\widehat{Y}(f)\right\rangle_n = Q(f) \sum^N_{k=1} e^{2\pi i f \hat{\Omega}_k \cdot \Delta \vec{x}_k /c} h_{0, k}^2(z_k) f^{-7/3} & \left[ \frac{\left(1+\cos^2\iota_k \right)^2}{4}  F_{i}'^{+}(k) F_{j}'^+(k) + \cos^2\iota_k F_{i}'^\times(k) F_{j}'^\times(k) + \right. \\
& \left.  \frac{\left(1+\cos^2\iota_k \right)}{2} \cos\iota_k \left(F_{i}'^+(k) F_{j}'^\times(k) + F_{i}'^\times(k) F_{j}'^+(k) \right) \right] ,
\end{split}
\end{align}

\noindent where

\begin{align}
F'^+ &= F^+ \cos 2\psi - F^\times \sin 2\psi, \\
F'^\times &= F^+ \sin 2\psi + F^+ \cos 2\psi.
\end{align}

\noindent Comparing Eq.~\ref{eq:Gamma} and Eq.~\ref{eq:<Y>_v4}, we observe that it is useful to define a {\em discrete overlap reduction function}, denoted $\Gamma_N(f)$, which encodes the signal-cancelling behaviour of $N$ discrete events

\begin{align}\label{eq:Gamma_n}
\begin{split}
\Gamma_N(f) \equiv \dfrac{1}{K_N} \sum^N_{k=1} e^{2\pi i f \hat{\Omega}_k \cdot \Delta \vec{x}_k /c} h_{0, k}^2(z_k) & \left[ \frac{\left(1+\cos^2\iota_k \right)^2}{4}  F_{i}^+(k) F_{j}^+(k) + \cos^2\iota_k F_{i}^\times(k) F_{j}^\times(k) + \right. \\
& \left.  \frac{\left(1+\cos^2\iota_k \right)}{2} \cos\iota_k \left(F_{i}^+(k) F_{j}^\times(k) + F_{i}^\times(k) F_{j}^+(k) \right) \right] ,
\end{split}
\end{align}

\noindent where $K_N$ is a normalisation factor that is averaged over all events

\begin{align}
K_N \equiv \sum_k h_{0,k}^2 (z) \left( \frac{\left(1+\cos^2\iota_k \right)^2}{4} + \cos^2\iota_k \right) .
\end{align}

\noindent We note that, by assumption, the $N$ events contributing to $\Gamma_N$ are too weak to be resolved, and so $\Gamma_N(f)$ is a theoretical quantity that we do not know from measurement.

As in section \ref{ensemble}, we can write Eq.~\ref{eq:<Y>_v4} in the form

\begin{equation}
  \left\langle\widehat{Y}(f)\right\rangle_n = Q(f) H_N(f) \Gamma_N(f) ,
\end{equation}

\noindent where $H_N(f) = K_N f^{-7/3}$ is the strain power spectral density for one realisation from a finite set of astrophysical sources.
(This expression for $H_N(f)$ is valid up to the gravitational-wave frequency of the last stable orbit, above which we assume $H_N(f)=0$.)
As before, we define

\begin{equation}\label{eq:gamma_n}
  \gamma_N(f) = (5/\sin^2\delta) \Gamma_N .
\end{equation}


\section{Numerical testing}\label{numerical}

This section is organised as follows. 
In subsection \ref{discreteORF}, we perform numerical simulations to qualitatively illustrate the behaviour of $\gamma_N(f)$ for different values of $N$. 
In subsection \ref{Ratio}, we calculate the bias that occurs when we search for a non-Gaussian astrophysical background with the estimator designed for a Gaussian background.
We also investigate how the results change if we include/exclude events loud enough to be detected individually.


\subsection{Simulation}\label{discreteORF}

Our numerical simulation uses the following model.
We consider a normally-distributed population of binary neutron stars with average mass $m_1=m_2=1.33M_\odot$ and width $\sigma_m = 0.03 M_\odot$.
This mass distribution takes into account both observational data of double pulsar systems \cite{Ozel12} as well as population synthesis models.  
We use a realistic redshift distribution which takes into account the star formation rate and delay time between the binary formation and coalescence~\cite{ETMDC1,Rates10}.
We assign random sky location using an isotropic distribution.
The cosine of the inclination angle $\cos\iota$ is chosen from a uniform distribution on $[-1,1]$.
The polarisation angle $\psi$ is chosen from a uniform distribution on $[0,2\pi]$.

We generate many realisations of the stochastic background, each with a fixed number of events $N$.
For each event, we calculate the matched filter signal-to-noise ratio $\rho$ in order to determine if it is loud enough to be individually detected:

\begin{align}
\rho^2 = \frac{5}{6} \frac{(GM_c(1+z))^{5/3} \Fcurl^2}{c^3 \pi^{4/3} d_L^2(z)} \int^{f_\text{LSO}} df \,  \frac{f^{-7/3}}{S_n(f)} .
\end{align}

\noindent Here $S_n(f)$ is the detector's strain noise power spectral density (taken to be the design sensitivity of Advanced LIGO), $f_\text{LSO}$ is the (redshifted) gravitational-wave frequency of the last stable orbit, and 

\begin{align}
  \Fcurl^2 \equiv \sum_i \left [ \frac{1}{4} (1+\cos^2 \iota)^2 (F_i'^{+})^2 + \cos^2 \iota (F_i'^{\times})^2 \right] 
\end{align}

\noindent characterises the network response.
The index $i$ runs over three detectors: LIGO Hanford, LIGO Livingston, and Virgo.
We exclude any events with $\rho \geq 8$.

For each realisation, we calculate $\gamma_N(f)$ (Eq. \ref{eq:gamma_n}) for the LIGO Hanford, LIGO Livingston detector pair. 
(The sensitivity contribution from the Virgo-LIGO pairs is small enough to ignore.) 
We carry out the calculation for different values of $N = 10^2, 10^3$...$10^6$.
In a $\unit[1]{yr}$-long dataset, $N\approx10^4$ corresponds to a pessimistic rate~\cite{Rates10} (see also Table~\ref{tab:Rates}) and so the (very pessimistic) values of $N = 10^2$ and $N = 10^3$ are included for pedagogic purposes.
The higher values of $N$ ($\sim10^4$--$10^6$) correspond to astrophysical rates ranging from pessimistic to realistic~\cite{Rates10}.
We do not include higher values of $N$ because, as we shall see, $N = 10^6$ events in one year of science data produce a signal which is already difficult to distinguish from a Gaussian background.

\begin{table}
\caption{\label{tab:Rates} A list of binary neutron star coalescence rate densities  as given in \cite{Rates10}. The first column labels whether a merger rate is optimistic ($R_\text{high}$), realistic ($R_\text{realistc}$) or pessimistic ($R_\text{low}$). The second column gives the rates of coalescing events per $\unit[]{Mpc^3}$ per $\unit[]{Myr}$. The third column gives the average time between successive events. The final column gives the total number of events in the universe that are expected to occur per year.
}
\begin{ruledtabular}
\begin{tabular}{ l c c c }
Expected Rate & $\dot{\rho}_0 (Mpc^{-3} Myr^{-1})$ & $\overline{\Delta t}$ (s) & $N_\text{events} yr^{-1}$ \\ \hline
$R_\text{high}$ & 10 & 1.25 & $2.5\times10^7$ \\ 
$R_\text{realistc}$ & 1 & 12.5 & $2.5\times10^6$ \\ 
$R_\text{medium-low}$ & 0.1 & 125 & $2.5\times10^5$ \\ 
$R_\text{low}$ & 0.01 & 1250 & $2.5\times10^4$ \\
\end{tabular}
\end{ruledtabular}
\label{tab:Rates}
\end{table}

\begin{figure*}
    \includegraphics[width=4in]{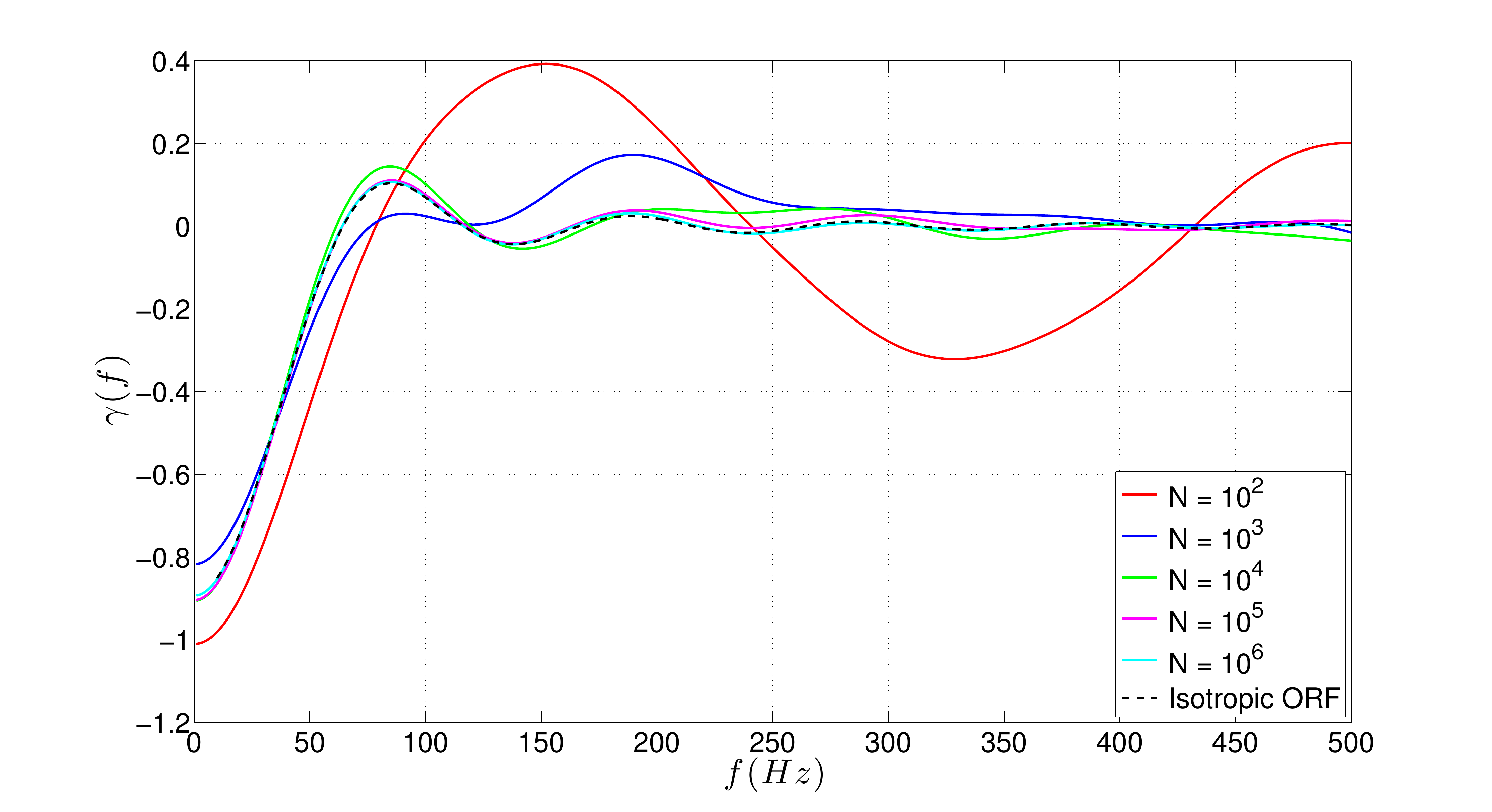}
    \caption{
      The discrete overlap reduction function $\gamma_N(f)$ for five different values of $N$.
      The overlap reduction function $\gamma(f)$ from a Gaussian isotropic background is shown by the black dashed line.
    }
    \label{fig:ORFs}
\end{figure*}

In Fig.~\ref{fig:ORFs}, we plot $\gamma_N(f)$ for individual realisations of the stochastic background, each with a different value of $N$.
For comparison, the standard overlap reduction function for an unpolarised, isotropic, Gaussian background $\gamma(f)$ is shown with a black line.
For small values of $N$, we see that $\gamma_N(f)$ can diverge significantly from $\gamma(f)$.
As $N$ increases, the overlap reduction function becomes closer to the Gaussian isotropic case.
Thus, Fig.~\ref{fig:ORFs} demonstrates how the discreteness of an astrophysical stochastic background can create spectral features, which are not expected for a Gaussian background.

\begin{figure*}
  \includegraphics[width=4in]{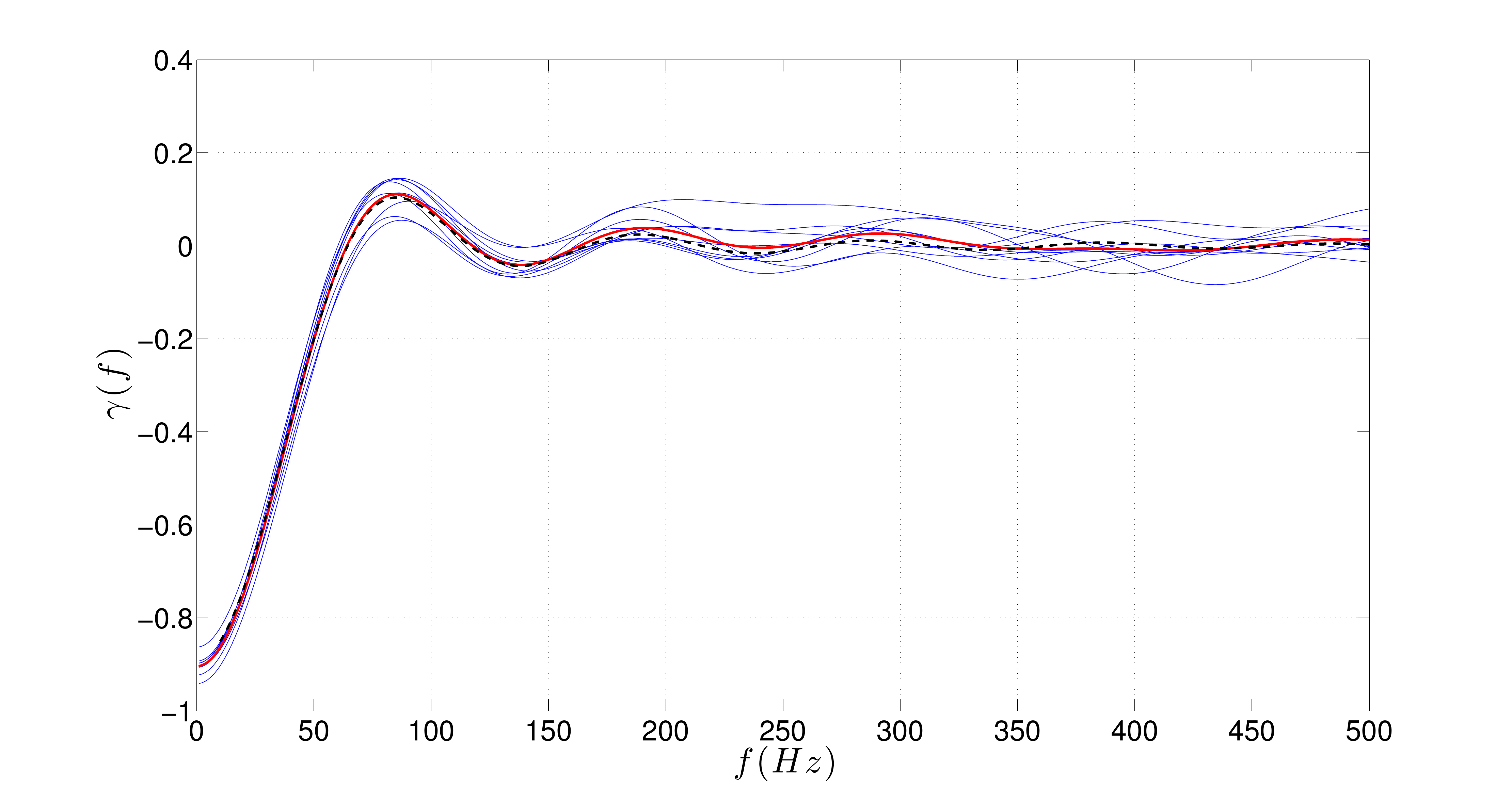}
  \includegraphics[width=4in]{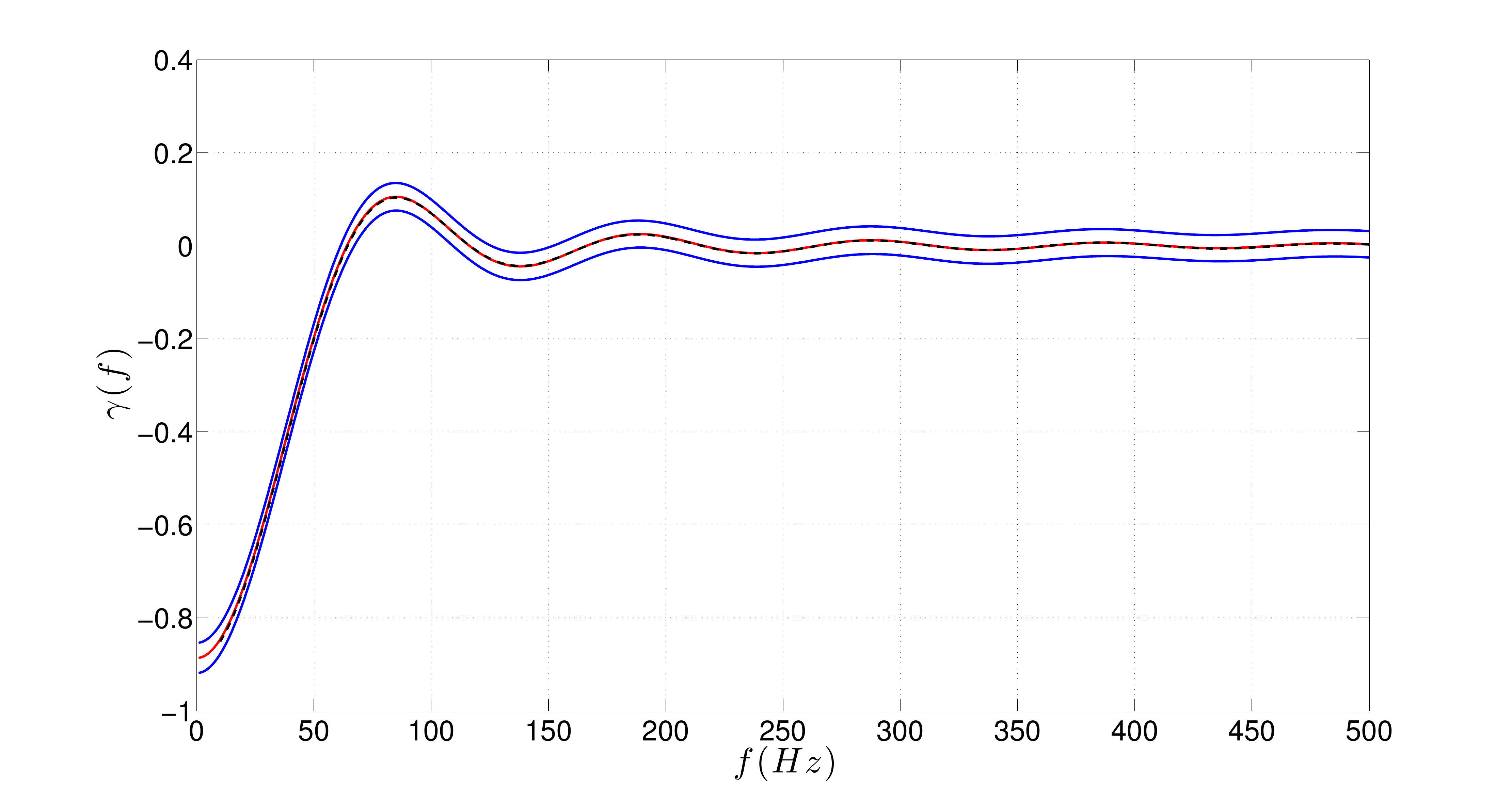}
  \includegraphics[width=4in]{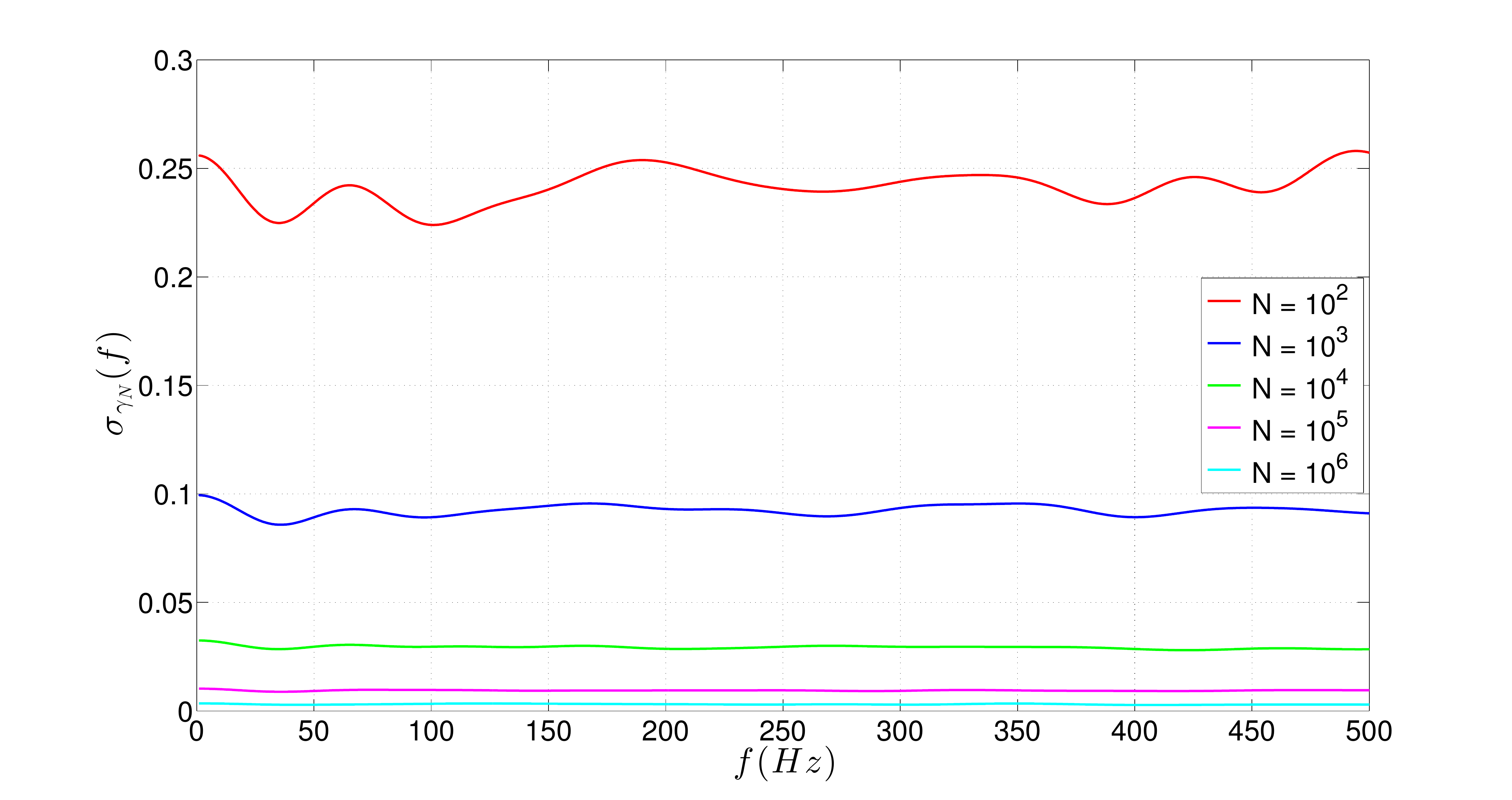}
  \caption{
    Top: ten realisations of the discrete overlap reduction function $\gamma_N(f)$ with $N=10^4$ events (blue).
    The mean of the blue curves is shown in red. The standard overlap reduction function $\gamma(f)$ is shown in dashed black.
    Middle: using a simulation of 1000 realisations of $N=10^4$ background sources, we calculate the standard deviation of $\gamma_N(f)$ at each frequency bin.
    The blue curves represent $\pm$ one standard deviation about the mean, which is shown in red.
    The dashed black corresponds to $\gamma(f)$.
    Bottom: variation in the overlap reduction function. We plot $\sigma_\gamma(f)$---the standard deviation of the discrete overlap reduction function as a function of frequency. 
    Each colour represents a different value of $N$. 
    The magnitude of $\sigma_\gamma(f)$ is approximately constant, which implies that the fractional error grows as $\gamma_N(f)$ becomes smaller at higher frequencies.
  }
\label{fig:ManyGammas}
\end{figure*}

In Fig.~\ref{fig:ManyGammas}a, we show ten realisations of $\gamma_N(f)$ for $N = 10^4$ (blue).
As one would expect, the mean of these ten realisations (red) is in good agreement with $\gamma(f)$ (black) as this can be considered as one realisation of $N = 10^5$ events.
By comparing the red and black traces, it is possible to get a qualitative sense of the typical fluctuations due to discreteness at a fixed value of $N$.
In Fig.~\ref{fig:ManyGammas}b, we plot $\gamma_N(f)\pm\sigma_\gamma(f)$ where $\sigma_\gamma(f)$ is the (numerically estimated) standard deviation of $\gamma_N(f)$ due to fluctuations arising from the discreteness of the background.
Finally, in  Fig.~\ref{fig:ManyGammas}c, we plot $\sigma_\gamma(f)$ to show that $\sigma_\gamma(f)$ is approximately constant in frequency.
Since $\gamma_N(f)$ tends to get smaller at higher frequencies, this implies that the fractional uncertainty $\sigma_\gamma(f)/\gamma_N(f)$ tends to become larger at higher frequencies.

\subsection{Bias}\label{Ratio}
By combining results for many independent frequency bins, it is possible to significantly increase the signal-to-noise ratio of a stochastic broadband search~\cite{locus}.
If the spectral shape of the stochastic background $\Omega_\text{gw}(f)$ is known, the expectation value (averaged over realisations of noise) of the optimal broadband estimator for an astrophysical background with discrete events is given by~\cite{Allen97}:

\begin{align} \label{eq:mean_u}
  \langle\widehat{Y}_\text{ISO}\rangle_n = \frac{3H_0^2}{20\pi^2} t_\text{obs} \int df \,  f^{-3} \Omega_\text{gw} \left (f \right ) \gamma_N \left ( f \right ) Q'(f) .
\end{align}

\noindent $Q'(f)$ is a filter function (not necessarily the same as $Q(f)$ for the narrowband estimator in Eq.~\ref{eq:Y(f)}) given by

\begin{align}\label{eq:Qprime}
  Q'(f) = \lambda \frac{\gamma \left ( f \right ) \Omega_\text{gw}(f)}{f^3 S_n\left ( f \right ) S_n \left ( f \right ) } .
\end{align}

\noindent Here, $\lambda$ is an overall normalisation constant and $\gamma(f)$ is the isotropic overlap reduction function. 
We have assumed, for the sake of simplicity, that the noise power spectral density $S_n(f)$ is the same for both detectors.
For the background of binary coalescences considered here, $\Omega_\text{gw}(f)\propto f^{2/3}$.

Substituting $Q'(f)$ into Eq. \ref{eq:mean_u}, we obtain

\begin{align}\label{eq:mismatch}
  \langle\widehat{Y}_\text{ISO}\rangle_n = \frac{3H_0^2}{20\pi^2} t_\text{obs} \int df \, \frac{\Omega_\text{gw}^2(f) \; \gamma_N \left ( f \right ) \gamma \left ( f \right )}{f^6 S_n(f) S_n(f) } .
\end{align}

\noindent We can think of Eq.~\ref{eq:mismatch} as the case where we apply an isotropic Gaussian filter $Q'(f)$ to an unknown background, which is in reality non-Gaussian.
If we had perfect knowledge of the $N$ events responsible for the observed background, we could calculate a more accurate estimator, $\hat{Y}_N$.
By the same line of reasoning, the (noise-averaged) expectation value of $\widehat{Y}_N$ in the presence of a known astrophysical background characterised by $N$ events is

\begin{align}
  \langle\widehat{Y}_N\rangle_n = \frac{3H_0^2}{20\pi^2} t_\text{obs} \int df \, \frac{\Omega_\text{gw}^2(f) \; \gamma^2_N \left ( f \right )  }{f^6 S_n(f) S_n(f) } .
\end{align}

\noindent By considering the ratio 

\begin{equation}\label{eq:ratio}
  R \equiv \frac{ \langle\widehat{Y}_N\rangle_n }{ \langle \widehat{Y}_\text{ISO}\rangle_n} ,
\end{equation}
we can characterise the fractional bias introduced into a stochastic search when we apply a Gaussian isotropic filter to a non-Gaussian background.

\begin{figure}
\centering
\begin{minipage}[h]{0.49\columnwidth}
\includegraphics[width=\columnwidth]{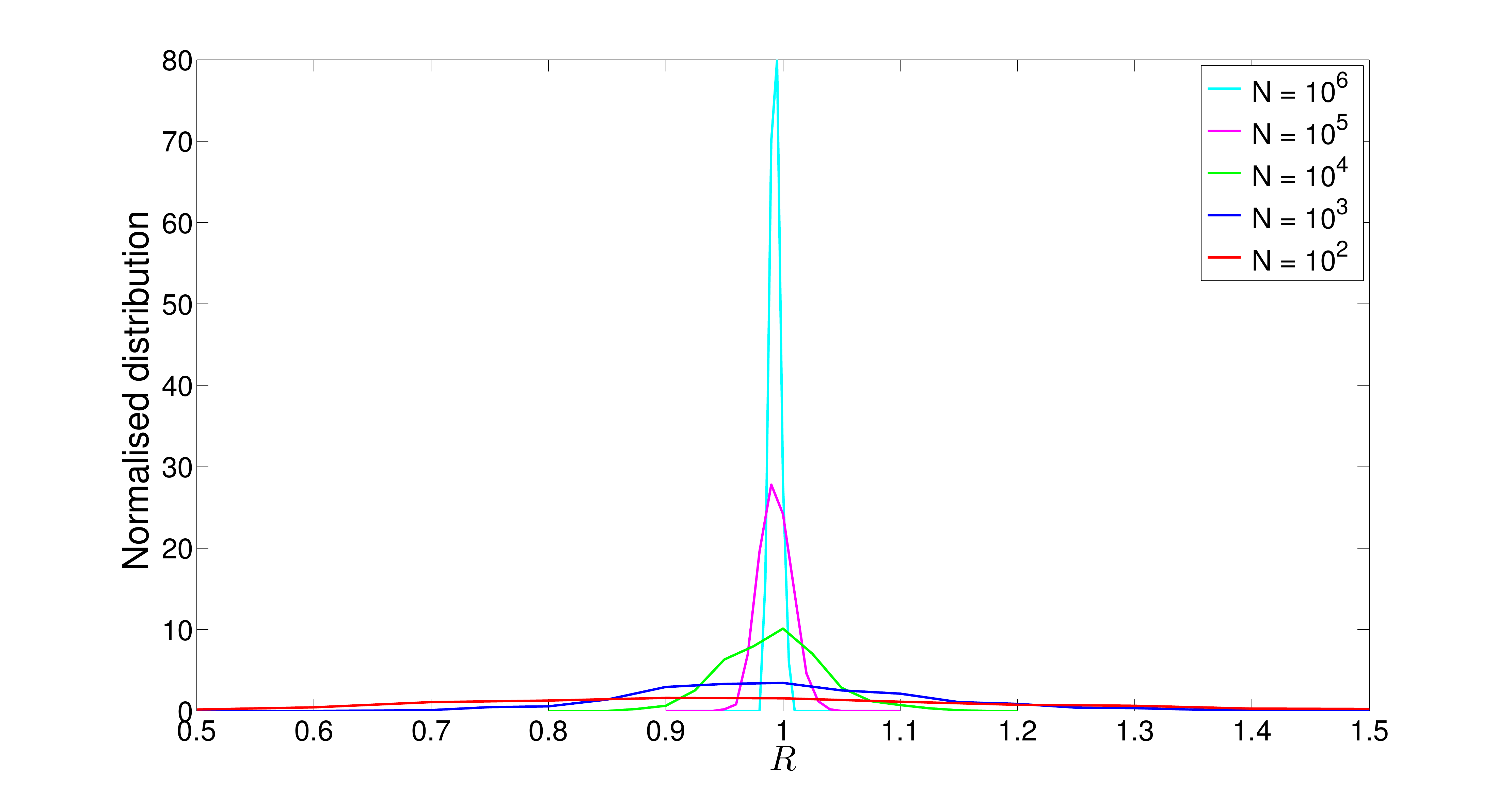}
\end{minipage}
\begin{minipage}[h]{0.49\columnwidth}
\includegraphics[width=\columnwidth]{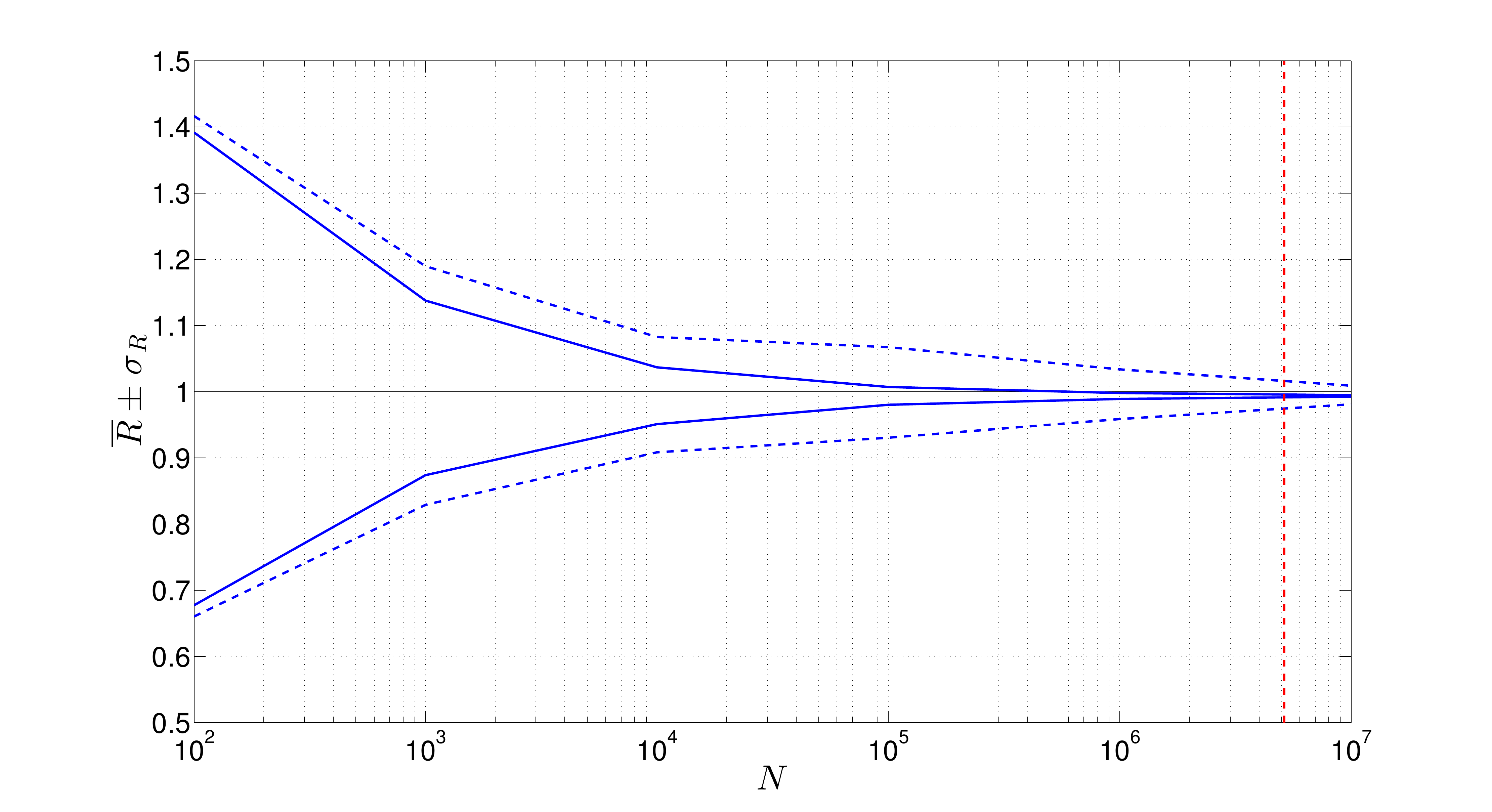}
\end{minipage}
\caption{
  Left: histogram of the fractional bias in a stochastic search $R$ (see Eq.~\ref{eq:ratio}) due to the discreteness of a non-Gaussian background.
  Each colour corresponds to a different value of $N$.
  Right: the standard deviation $\sigma_R$ of the left-hand-side histograms as a function of $N$ (blue).
  The dashed lines include all events whereas the solid lines exclude events loud enough to resolve individually with matched filtering.
  The dashed red line corresponds to the number of events required to produce a stochastic signal that can be detected with $\text{SNR} = 2$ (see Eq.~\ref{eq:stochSNR}) using Advanced LIGO with an observational period of $t_\text{obs}=\unit[1]{yr}$.
  Note that $\sigma_R$ does not depend on $t_\text{obs}$ whereas $\text{SNR}\propto t_\text{obs}^{1/2}$.
}
\label{fig:Ratios}
\end{figure}

In Fig.~\ref{fig:Ratios}a, we show histograms of $R$ for different values of $N$.
As $N$ increases, the width of the distribution of $R$ decreases, indicating that the fractional bias decreases as expected.
In  Fig.~\ref{fig:Ratios}b, we plot the standard deviation of the distribution of $R$ as a function of $N$.
The dashed red line indicated the number of events that are required to occur within an observational period of $t_\text{obs} = \unit[1]{yr}$ in order to obtain a stochastic signal-to-noise ratio of 2.
The average signal-to-noise ratio of a stochastic search is given by \cite{Allen97}

\begin{equation}\label{eq:stochSNR}
  \text{SNR} \approx \frac{3 H_0^2}{10 \pi^2} \sqrt{t_\text{obs}} \left[ \int df \frac{\Omega^2_\text{gw}(f) \gamma^{2}(f) }{ f^6 S_n(f) S_n(f)} \right ]^{1/2} .
\end{equation} 

Also in Fig.~\ref{fig:Ratios}b, we show how the results change if we do not remove individually detectable events with $\rho\geq8$; see the dashed blue lines.
We find that the inclusion of loud events changes standard deviation of the fractional bias $R$ by $\lesssim8\%$ depending on the value of $N$.

\section{Conclusions}\label{conclusions}

Many previous studies of the stochastic gravitational-wave background have assumed a signal that is isotropic, unpolarised, and Gaussian.
However, non-Gaussian backgrounds from compact binary coalescence represent one of the most exciting sources for second-generation detectors such as Advanced LIGO and Virgo.
In this paper, we investigated the statistical properties of stochastic backgrounds originating from a discrete set of astrophysical sources and how they will appear in future cross-correlation searches.
In the course of our investigation, we found it useful to define a novel description of astrophysical backgrounds: a discrete overlap reduction function.
We find that the discreteness of astrophysical backgrounds is unlikely to produce a measurable bias in upcoming observations by second-generation detectors.

Here we focused on upcoming advanced detectors observing a population of binary neutron star sources.
However, we note that the situation may be more complicated for the proposed third-generation Einstein Telescope \cite{ETMDC1}.
In particular, we raise the possibility that the removal of above-threshold binary events may create a selection bias.
This is because we expect face-on events, directly above the detector, will be preferentially detected compared to events with less favourable orientations and locations, which in turn, may create an apparent anisotropy.
The effect may be more pronounced for the Einstein telescope (with only one detector) versus a network of 2--5 advanced detectors.
Future work will characterise the magnitude of this effect for the Einstein Telescope.


\begin{acknowledgments}
DM acknowledges the PhD financial support from the Observatoire de la C$\hat{o}$te d'Azur and the PACA region.
DM would also like to thank the LIGO Laboratory and the California Institute of Technology for a visiting scientist fellowship under which part of this work was conducted.
ET is a member of the LIGO Laboratory, supported by funding from United States National Science Foundation.
LIGO was constructed by the California Institute of Technology and Massachusetts Institute of Technology with funding from the National Science Foundation and operates under cooperative agreement PHY-0757058.
This paper has been assigned document number LIGO-P1400019.
\end{acknowledgments}

\bibliography{DiscreteORF_v5}

\end{document}